\documentstyle[12pt,fleqn]{article}
\def\beq{\begin{equation}}
\def\eeq{\end{equation}}
\def\bA{{\bf A}}
\def\1{\mbox{\small1\hskip-0.35em\normalsize1}}
\def\6{\langle }
\def\9{\rangle }

\begin{document}

\renewcommand{\thefootnote}{\fnsymbol{footnote}}

\vspace*{10mm}
\begin{center}
{\large {\bf Convex probability domain\\[1mm] of generalized quantum
measurements}}\\[15mm]

Asher Peres\footnote{Electronic address: peres@photon.technion.ac.il}
and Daniel R. Terno\footnote{Electronic address:
terno@physics.technion.ac.il} \\[8mm]
{\sl Department of Physics, Technion---Israel Institute of Technology,
32\,000 Haifa, Israel}\\[15mm]

{\bf Abstract} \end{center}

\noindent Generalized quantum measurements with $N$ distinct outcomes
are used for determining the density matrix, of order $d$, of an ensemble
of quantum systems. The resulting probabilities are represented by a
point in an $N$-dimensional space. It is shown that this point lies
in a convex domain having at most $d^2-1$ dimensions.\newpage

\noindent In elementary quantum measurement theory, a test performed on
a quantum system is represented by a complete set of orthogonal
projection operators ${\bf P}_m$, where the label $m$ takes at most $d$
different values ($d$ is the dimensionality of the Hilbert space,
assumed finite). The probability of obtaining outcome $m$ of that test,
following the preparation of a quantum ensemble in a state $\rho$, is
$p_m={\rm tr}\,(\rho{\bf P}_m)$. If $\rho$ is arbitrary, the only
constraint on these probabilities is $\sum_m p_m=1$.

It is well known that this type of test is not optimal if only a finite
number of quantum systems can be observed. (As a concrete example, we
receive five photons from a distant source, and we want a good estimate
of their polarization. What is the best strategy?) In such a case, more
information may be derived from a {\it positive operator valued
measure\/} (POVM) [1,~2] with $N>d$ different outcomes. Such a POVM is a
set of $N$ positive matrices $\bA_\mu$, which in general do not commute,
but still satisfy $\sum_\mu \bA_\mu=\1$, where \1\ is the unit matrix in
$d$ dimensions, and $\mu$ is an arbitrary label running from 1 to $N$.
If the quantum system is prepared in state $\rho$, the probability to
get outcome $\mu$ is

\beq p_\mu(\rho) \equiv p(\bA_\mu|\rho)={\rm tr}\,(\rho\bA_\mu).
\label{pd} \eeq

To each preparation $\rho$ of the system, we thus associate $N$
probabilities, $p_1,\ p_2,\ldots\ ,p_N$. We refer to this set of
positive numbers as a point $P(\rho)$ in probability space. The set of
all output points is labelled by $P$. Density matrices form a convex
set whose extreme points are pure states~[3]. The linear
relation~(\ref{pd}) between input states and output probabilities
implies that the set of points $P$ is also convex:

\beq P(\rho)\equiv P(x \rho_1 +(1-x) \rho_2)
  =x P(\rho_1)+(1-x) P(\rho_2). \eeq 
Thus the shape of the hypersurface that bounds the domain of the points
$P(\rho)$, for all possible preparations of the system, is determined by
the outputs for the pure states only.

Obviously $\sum_\mu p_\mu=1$, so that the points $P(\rho)$ lie on a
hyperplane of dimension \mbox{$(N-1)$.} However, the results of
generalized measurements are subject to stronger constraints (which may
be important for the statistical analysis of experimental results). Let
$D$ be the number of linearly independent parameters in $\rho$ (for a
generic density matrix in a $d$-dimensional complex Hilbert space,
$D=d^2-1$). The following proposition will now be proved: {\it If
$N>(D+1)$, the output of any POVM is confined to a D-dimensional
subspace.}

Indeed, let us write the elements of a generic density matrix in terms
of real (symmetric) and imaginary (anti\-symmetric) parts,

\beq \rho_{mn}=\xi_{mn}+i\eta_{mn}. \eeq
There are $d(d-1)/2$ independent elements $\eta_{mn}$ and
$(d+2)(d-1)/2$ independent $\xi_{mn}$, because of the condition $\rm
tr\,\rho=1$ which can be written

\beq \xi_{dd}=1-\sum_{n=1}^{d-1} \xi_{nn}. \eeq
Likewise, the elements of each POVM matrix $\bA_\mu$, of order $d$, can
be written as $x^\mu_{mn}+iy^\mu_{mn}$ in terms of $d^2$  real
parameters. We thus obtain from Eq.~(\ref{pd}),

\beq p_\mu(\rho)=\sum_{m=1}^{d-1}(x_{mm}^\mu-x_{dd}^\mu)\,\xi_{mm}+
 2\sum_{m=1}^d\sum_{n>m} (x^\mu_{mn}\xi_{mn}+y^\mu_{mn}\eta_{mn})
 +x_{dd}^\mu.  \eeq 

Thus $P(\rho)$ is obtained from $\rho$ by an affine transformation~[4]

\beq {\bf p=Mr+c}, \label{trans}\eeq
where {\bf p} is a `vector' consisting of any $N-1$ components $p_\mu$
(the remaining component is obtained from $\sum_\mu p_\mu=1$). Likewise
{\bf r} is a vector of $D$ linearly independent parameters of $\rho$.
The matrix {\bf M}, with $N-1$ rows and $D$ columns, depends only on the
POVM used for the test; and {\bf c} is a vector whose $N-1$ components
are $x_{dd}^\mu$, which also are parameters of the POVM.  Explicitly,
the $\mu$-th row of {\bf M}, which is

\beq (x^\mu_{11}-x^\mu_{dd}) \ \ldots \ (x^\mu_{d-1,d-1}-x^\mu_{dd})
 \quad 2x^\mu_{12} \ \ldots \ 2x^\mu_{d-1,d} \quad
 2y^\mu_{12} \ \ldots \ 2y^\mu_{d-1,d}, \eeq
and 

\beq {\bf r}^T=(\xi_{11} \ \ldots \ \xi_{d-1,d-1} \quad \xi_{12} \
\ldots \ \xi_{d-1,d} \quad \eta_{12} \ \ldots \ \eta_{d-1,d}),\eeq
have $D=d^2-1$ real components.

If $N-1>D$, the rank of {\bf M} is at most $D$, and any $D+1$ vectors
$\bf \tilde{p}=Mr$ are linearly dependent. A translation by the constant
vector {\bf c} in the $(N-1)$-dimensional vector space transforms a
$D$-dimensional subspace into another $D$-dimensional subspace. Thus the
output of any POVM on a system whose density matrix has $D$ linearly
independent parameters is confined to a $D$-dimensional subspace of the
probability space.

Next, let us examine the shape of the surface that encloses the domain
of $P(\rho)$. The set of density operators, and therefore the set of
probabilities, are convex. The extreme points of these sets are the pure
states, which are defined by $2(d-1)$ real parameters, and the
probabilities corresponding to these pure states, respectively. Thus any
interior point of the $D$-dimensional set $P$ is a convex combination of
the extreme points of that set, which lie on a $2(d-1)$-dimensional
hypercurve.

Note that any density matrix $\rho$ of rank $d$ can be written as a
convex combination of no more than $d$ pure density matrices,
corresponding to the eigenvectors of $\rho$. As a result, any interior
point of $P$ can be obtained from at most $d$ extreme points. This
result ought to be compared with Caratheodory's theorem~[4], which
states that any interior point of an arbitrary convex set of dimension
$D$ can be obtained as a convex combination of $D+1$ (or fewer) extreme
points of that set. Here, $D+1=d^2$. The smaller number of extreme
points needed in the present case is due to the fact that density
matrices are not an {\it arbitrary\/} convex set (they are positive and
have unit trace).

As a simple example, consider the case of spin-$1\over2$ systems. Their
states can be described by means of a Bloch sphere. The pure states
correspond to points on the surface of the sphere, and mixed states lie
in its interior. With our notations, we have

\beq \rho=\left( \begin{array}{ccc} x_{11} & & x_{12}+iy_{12} \\
 x_{12}-iy_{12} & & 1-x_{11} \end{array}\right),\eeq
where the three parameters are subject to the positivity condition

\beq x_{11}\,(1-x_{11})-x_{12}^2-y_{12}^2\ge0 \label{blo}\eeq 
The transformation (\ref{trans}) is linear. Therefore the Bloch sphere
is transformed into another quadratic surface, usually an ellipsoid.
Exceptionally, if a POVM element has unit norm (so that the
corresponding $p_\mu$ can be equal to 1, and then all the other $p_\mu$
vanish), we have a cone.

In particular, consider a POVM with four elements,
$\bA_\mu=(\1+{\bf a}_\mu\cdot\mbox{\boldmath$\sigma$})/4$, where
the four unit vectors ${\bf a}_\mu$ form a regular tetrahedron in a real
3-dimensional Euclidean space, and {\boldmath$\sigma$} denotes the three
Pauli matrices. Likewise, any state $\rho$ can be written as
$\rho=(\1+{\bf n}\cdot\mbox{\boldmath$\sigma$})/2$. We thus have

\beq p_\mu={\rm tr}\,(\rho\bA_\mu)=(\1+{\bf a_\mu\cdot n})/4,\eeq
whence

\beq \sum_{\mu=1}^4(p_\mu-\mbox{$1\over4$})^2={\bf n}^2/12.\eeq
The Bloch sphere is thus mapped into a 3-dimensional sphere of radius
$1/\sqrt{12}$, centered at $p_\mu={1\over4}$\,, and lying in the
hyper\-plane $\sum_\mu p_\mu=1$. If we want to parametrize that
hyper\-plane with three of the $p_\mu$, we substitute in the above
equation $p_4=1-p_1-p_2-p_3$. We then obtain an ellipsoid in a
3-dimensional space, as shown in Fig.~1. It is also possible to use as
coordinates suitable linear combinations of the $p_\mu$, orthogonal to
$\sum_\mu p_\mu$, such as
\beq \begin{array}{l} x=p_1+p_2-p_3-p_4,\\
  y=p_1-p_2+p_3-p_4,\\ z=p_1-p_2-p_3+p_4.\end{array}\eeq
The Bloch sphere is then mapped into a sphere $x^2+y^2+z^2\le{1\over3}$.

The case of spin-1 systems is more complicated. A generic density matrix
can be written in terms of its eigenstates as

\beq \rho=\sum_{j=1}^3 \lambda_j |v_j\9\6v_j|. \label{eigenv} \eeq
This is as a convex combination of three extreme points. Any pure state,
such as the above eigenstates, can be parametrized, with a suitable
choice of its phase, as

\beq |v\9=(\sin\theta \cos\phi\, e^{i\alpha},
\sin\theta \sin\phi\, e^{i\beta},\cos\theta), \eeq 
where

\beq 0\leq\theta,\ \phi\leq\pi/2,\qquad\qquad{\rm and}\qquad\qquad
0\leq\alpha,\ \beta<2\pi.\eeq
All the components of the corresponding pure $\rho$, which is a matrix
of rank~1, are functions of the four parameters $\theta,\ \phi,\
\alpha$, and $\beta$.  Thus all the probabilities $p_\mu={\rm tr}
\,(\rho\bA_\mu)$ are also functions of these four angles. This gives the
extreme points of the set $P$: they form a four-parameter hypersurface
in an eight-dimensional space $S$ (which is itself embedded in the
$N$-dimensional space of the $p_\mu$). The rest of the boundary of $S$,
corresponding to density matrices of rank two, lies on the segments
between any pair of extreme points. The interior points of $S$ can be
obtained by a convex combination of three suitably chosen extreme
points, as in Eq.~(\ref{eigenv}). All these considerations are readily
extended to quantum systems whose Hilbert spaces have more than three
dimensions: there are $(d-1)$ polar angles like $\theta$ and $\phi$, and
$(d-1)$ phases like $\alpha$ and $\beta$.

Finally, let us consider potential applications of the above results to
the analysis of experimental data. The probabilities $p_\mu$ cannot be
measured exactly, as this would require testing an infinite number of
quantum sytems. If only $n$ systems are available, and the $\mu$-th
outcome is found to occur $n_\mu$ times (so that the experimenter
records a set of $N$ integers or zeros), then the $N$ ratios
$q_\mu=n_\mu/n$ are the only data available for evaluating the true
$p_\mu$. Obviously, $\sum_\mu q_\mu=1$, just like $\sum_\mu p_\mu$, but
the other constraints on $p_\mu$ may not be satisfied. In particular, if
$N>D+1$, the point $Q=\{q_\mu\}$ will not in general lie in the
hyperplane of dimension $D$ to which the point $P$ is restricted.

How far can $Q$ be from the true $P$? Each one of the experimental data
$n_\mu$ has an expected binomial distribution with dispersion

\beq \Delta n_\mu=[n\,p_\mu(1-p_\mu)]^{1/2}\simeq
  [n_\mu(n-n_\mu)/n]^{1/2},\eeq 
where the last expression is valid if $n_\mu\gg1$. We can imagine an
error box with sides equal to $\Delta q_\mu$, centered at the point $Q$,
and we then have to examine where that error box overlaps with the
hyperplane to which $P$ is constrained.

Obviously, it is best to design the experiment so as to have
$N=D+1=d^2$, and not more than that. A larger value of $N$ leads to a
less efficient use of the experimental data. This result is reminiscent
of Davies's theorem~[5] which deals with a related question, namely how
to maximize the mutual information obtainable from a set of
non-orthogonal signals. The theorem asserts that no more than $d^2$
outputs are needed.

If $N=D+1$, the only question is whether $Q$ lies in the convex domain
of $P$, namely whether the resulting $\rho$, obtained by solving
Eq.~(\ref{pd}), is a positive matrix. If it is, then $Q$ is the best
estimate for the true $P$. In the opposite case, we may consider whether
the error bars $\Delta q_\mu$ reach the convex domain of $P$. If they
do, we still get a reasonable estimate. If even this fails, the
experimenter has to test a larger number of samples in order to have a
meaningful measurement.

\bigskip\noindent{\bf Acknowledgments}\medskip

DRT was supported by a grant from the Technion Graduate School. Work by
AP was supported by the Gerard Swope Fund, and the Fund for
Encouragement of Research.\bigskip  \clearpage

\noindent{\bf References}

\frenchspacing \begin{enumerate}
\item Helstrom C W 1976 {\it Quantum Detection and Estimation Theory}
(New York: Academic Press) 
\item Peres A 1993 {\it Quantum Theory: Concepts and Methods\/}
(Dordrecht: Kluwer)
\item Davies E B 1976 {\it Quantum Theory of Open Systems\/} 
(New York: Academic Press)
\item Kelly P J and Weiss M L 1979 {\it Geometry and Convexity\/}
(New York: Wiley)
\item Davies E B 1978 {\it IEEE Trans. Inform. Theory\/} {\bf IT-24} 239
\end{enumerate}\vfill

\noindent FIG. 1. \ The probability ellipsoid that corresponds to the
Bloch sphere is tangent to the plane $p_1+p_2+p_3=1$ (that is $p_4=0$)
at the point $p_1=p_2=p_3={1\over3}$ and likewise it is tangent to each
coordinate plane at the point where the two coordinates in that plane
are~$1\over3$.
\end{document}